\documentclass[a4paper,fleqn,usenatbib]{mnras}
\usepackage{graphicx}
\usepackage{amssymb}
\usepackage{amsmath}
\usepackage{booktabs}
\usepackage{varwidth}
\usepackage[T1]{fontenc}
\usepackage{ae,aecompl}
\usepackage{natbib,times}
\usepackage{lscape}
\usepackage{array}
\usepackage{setspace}
\usepackage{helvet}
\usepackage{float}
\usepackage{graphicx}
\usepackage{subcaption}
\usepackage{multirow}
\newcommand{\PreserveBackslash}[1]{\let\temp=\\#1\let\\=\temp}
\newcolumntype{C}[1]{>{\PreserveBackslash\centering}p{#1}}
\newcolumntype{R}[1]{>{\PreserveBackslash\raggedleft}p{#1}}
\newcolumntype{L}[1]{>{\PreserveBackslash\raggedright}p{#1}}

\begin{document}

\title[The XMM-Newton Galaxy Cluster sample] {A sample of 3403 galaxy
  clusters identified in XMM-Newton X-ray images}

\author[Z. S. Yuan et al.]
       {Z. S. Yuan,$^{1,2}$ \thanks{E-mail: zsyuan@nao.cas.cn}
        Z. L. Wen,$^{1,2}$
        W. Xu$^{1,3}$ 
        and J. L. Han,$^{1,2,4}$
\\
1. National Astronomical Observatories, Chinese Academy of Sciences, 
           20A Datun Road, Chaoyang District, Beijing 100101, China\\
2. CAS Key Laboratory of FAST, NAOC, Chinese Academy of Sciences,
           Beijing 100101, China\\
3. School of Physics and Astronomy, 
           Beijing Normal University, Beijing 100875, China\\
4. School of Astronomy, University of Chinese Academy of Sciences,
           Beijing 100049, China}

\date{Accepted XXX. Received YYY; in original form ZZZ}

\label{firstpage}
\pagerange{\pageref{firstpage}--\pageref{lastpage}}
\maketitle


\begin{abstract}
Currently, the number of galaxy clusters identified using galaxy data
has far exceeded the number derived from intracluster medium data. In
this study, we used positional information from large optical cluster
catalogues to search for previously unrecognized X-ray galaxy clusters
in archival XMM-Newton data. We successfully identified 1490 galaxy
clusters in X-ray images for the first time. By incorporating 1913
previously known X-ray clusters, we constructed a sample of 3403
galaxy clusters observed by XMM-Newton. Our cluster mass estimates
show broad consistency with previous measurements. Comparative
analyses between the known and newly identified subsamples revealed
that new X-ray clusters exhibit systematically higher redshifts, lower
masses, and smaller X-ray-to-optical mass ratios, but show no
systematic differences in dynamical properties. The newly identified
X-ray clusters are a valuable addition to previous X-ray samples and
are important for future statistical studies.
\end{abstract}


\section{Introduction}
\label{intro}

Galaxy clusters are the largest gravitationally bound systems in the
Universe and play important roles in studies spanning galaxy evolution
\citep[e.g.,][]{doc+97,bfs22}, cluster physics
\citep[e.g.,][]{zmj10,bbp+12,bj14}, and cosmology
\citep[e.g.,][]{whl10,hhw+12,bcc14}. Their mass budget comprises three
primary components: dark matter ($\sim$80\% of the total mass), the
intracluster medium (ICM, $\sim$15-17\%), and member galaxies
($\sim$3-5\%). Galaxy clusters typically host dozens to thousands of
member galaxies observable in the optical and infrared bands
\citep[e.g.,][]{yzj+03,fcc+12}. The ICM, the dominant baryonic matter
within galaxy clusters, is observable via thermal bremsstrahlung
emission in X-rays \citep[e.g.,][]{s86,bw10} and the
Sunyaev-Zel'dovich (SZ) effect -- produced by inverse Compton
scattering of cosmic microwave background (CMB) photons at millimeter
wavelengths \citep[e.g.,][]{sz72,b99}. For the elusive dark matter
component, its distribution and properties can be probed through
gravitational lensing of background galaxies
\citep[e.g.,][]{tvw90,cbg+06}.

The X-ray emission from the ICM is intrinsically coupled to the
gravitational potential of galaxy clusters, rendering it relatively
unaffected by the projection effect and hosting a robust correlation
with the total cluster mass. In the past, researchers have estimated
cluster masses based on their X-ray emissions for complete samples,
and further constrained the cosmological parameters with their mass
distribution \citep[e.g.,][]{bcc14,gpv+22}. However, observational
studies have revealed a statistically significant discrepancies
between the cosmological constant inferred from X-ray-selected cluster
samples and values derived from CMB measurements, a critical issue
termed the "$\sigma_8$ tension" \citep[see][and references
  therein]{pab+19}. A plausible solution to this tension lies in
overestimations of completeness in prior X-ray cluster samples,
specifically the possibility that a non-negligible fraction of
clusters may have been undetected due to observational
limitations. For instance, recent studies have identified galaxy
clusters where the hot gas has been stripped away due to processes
like violent merger events \citep[e.g.,][]{why24}, or clusters with
exceptionally flat X-ray surface brightness profiles that difficult to
be distinguished from background noise \citep[e.g.,][]{xrp+22}.

Previously, the strategy for identifying galaxy clusters in X-ray
images typically involved first selecting extended X-ray sources as
cluster candidates, followed by searching for optical counterparts for
verification and determining their redshifts
\citep[e.g.,][]{kcs+21,blk+24,kcl+24}. This traditional workflow
yields samples of galaxy clusters that follow a specific selection
function, which is derived from mock data. In practice, however, some
detectable clusters are not identified by traditional methods. For
example, in X-ray surveys with limited angular resolution, some galaxy
clusters may appear as point sources and thus be missed during cluster
identification \citep[e.g.,][]{blp+22}. Furthermore, to increase the
subsequent confirmation efficiency, threshold criteria (e.g., flux
limits, source extent parameters, or minimum signal-to-noise ratios)
are often applied to filter cluster candidates
\citep[e.g.,][]{eeb+98,eea+00,kcs+21}. These thresholds unavoidably
lead to the omission of some faint X-ray galaxy clusters. Studies have
also revealed that cluster samples constructed with X-ray data may be
biased toward containing more relaxed clusters, as relaxed systems,
compared to merging clusters, possess bright cool-cores that are
easier to detect in X-ray images \citep[e.g.,][]{wh13}.

The number of galaxy clusters identified in X-ray data has surpassed
ten thousands \citep{blk+24}, whereas over one million galaxy clusters
have been identified through optical surveys
\citep[e.g.,][]{zgx+21,yxh+21,wh24}. When a galaxy cluster is already
been identified in the optical band, its X-ray emission can be
directly probed at the corresponding location in X-ray images. This
approach not only enhances the accuracy of X-ray emission detection
but also enables relaxation or elimination of the threshold criteria
traditionally applied in X-ray cluster candidate selection. As a
result, this method reduces sample selection biases and may
substantially increase the detection rate of X-ray clusters. Using
positions from optical cluster catalogues \citep[e.g.,][]{wh15}, we
identified 49 previously unrecognized X-ray clusters in Chandra images
\citep{yh20} and 416 newly-confirmed X-ray clusters in XMM-Newton
images \citep{yhw22}.

Recently, \citet[][hereafter WH24]{wh24} constructed the largest
galaxy cluster catalogue using Data Release 10 (DR10) of the Dark
Energy Spectroscopic Instrument (DESI) Legacy Imaging Surveys
\citep{dsl+19}, which provide both extensive sky coverage and deep
imaging depth. This catalogue contains 1.58 million galaxy clusters
and over half of them are identified for the first time. Furthermore,
\citet[][hereafter WHY18]{why18} constructed a galaxy cluster sample
by incorporating data from all-sky surveys, including the Two Micron
All Sky Survey (2MASS), Wide-field Infrared Survey Explorer (WISE) and
SuperCOSMOS. In this work, we aim to identify both known and
previously unrecognized X-ray clusters in archival XMM-Newton data by
using existing X-ray \citep{kcs+21,blk+24,sap+24} and optical
\citep{why18,wh24} cluster catalogues. We will further investigate
whether the properties of newly-identified X-ray clusters differ
systematically from those of previously known ones, providing critical
insights to inform future statistical studies.

This paper is structured as follows. In Section~\ref{sect2}, we
describe the data collection and processing procedures, the method for
identifying previously known and unrecognized X-ray clusters, and the
estimation of cluster parameters. In Section~\ref{sect3}, we present
the statistical properties of our sample. We conclude with a summary
of key findings in Section~\ref{sect4}. Throughout this work, we adopt
a flat $\Lambda$CDM cosmology with $H_0=70~\rm km~s^{-1}~Mpc^{-1}$,
$\Omega_{m}=0.3$ and $\Omega_{\Lambda}=0.7$.

\section{Data processing and cluster identification}
\label{sect2}
\subsection{Data collection and processing}
\label{data}

We retrive X-ray data from the XMM-Newton Science Archive (XSA)
repository\footnote{https://nxsa.esac.esa.int/nxsa-web/\#search}. Observations
are selected from the XSA system with the following criteria. (1) The
\textsl{Duration} is set to ``$\ge10,000$'', corresponding to an
on-time threshold of 10 ks. (2) The observation \textsl{Availability}
is set to ``\textsc{Public}''. (3) The observation \textsl{Status}
marked as ``\textsc{Processed}''. (4) For the \textsl{Instrument
  Configuration}, any of the CCD systems, i.e., MOS1, MOS2 and EPN,
are adopted. For MOS1/MOS2 systems, the \textsl{Instrument Mode} is
set as ``\textsc{Full Frame}'' and the \textsl{Filters} is set as
``\textsc{Any}''; for the EPN system, the \textsl{Instrument Mode} is
set as ``\textsc{Extended Full Frame}'' or ``\textsc{Full Frame}'' and
the \textsl{Filters} is set as ``\textsc{Any}''. (5) To search as many
X-ray galaxy clusters as possible, all the \textsl{Proposal Category}
are selected, and all the \textsl{Proposal Program} are used except
the ``\textsc{Calibration}''. (6) Furthermore, only ``\textsc{Pointed
  Observations}'' and ``\textsc{EPIC Exposures}'' are selected. This
query yielded 10,477 observations retrieved from the XSA system as of
May 2025. Among these observations, many of them cover similar sky
regions. For duplicated observations, we generally retain the one with
maximal exposure time. Finally, we obtain 7,075 archival observations
from the XSA repository.

The X-ray data were processed in the XMM-Newton Science Analysis
System \citep[SAS, version 21.0.0,][]{gdf+04} with the latest Current
Calibration Files (CCFs). We downloaded Observation Data Files (ODFs)
from the XSA repository and processed them following the official
tutorial for extended
sources\footnote{https://www.cosmos.esa.int/web/xmm-newton/sas-thread-esasimage},
refering the section 2.2 of \citet{yhw22} for details. Nevertheless,
there are some modifications in this work compared to
\citet{yhw22}. First, the energy range used in this work is [0.5 -
  2.0] keV, rather than [0.5 - 5.0] keV adopted in the \citet{yhw22},
to ensure consistency with recent X-ray flux or luminosity estimates
\citep[e.g.,][]{kcs+21,blk+24}. Second, some tools in the latest SAS
version 21.0.0 are enhanced comparing to those used in
\citet[][version 18.0.0]{yhw22}. For example, we use the new command
\texttt{combimage}, rather than the \texttt{comb} command in version
18.0.0, to combine different exposure images observed by the MOS1,
MOS2 and EPN to enhance the signal-to-noise ratio of the images. The
new task \texttt{binadapt}, rather than the \texttt{adapt} task in
older version, is used to correct the exposure time and produce the
count rate image directly.

\begin{figure*}
\centering
\begin{subfigure}{0.7\textwidth}
\includegraphics[width=\textwidth, angle=0]{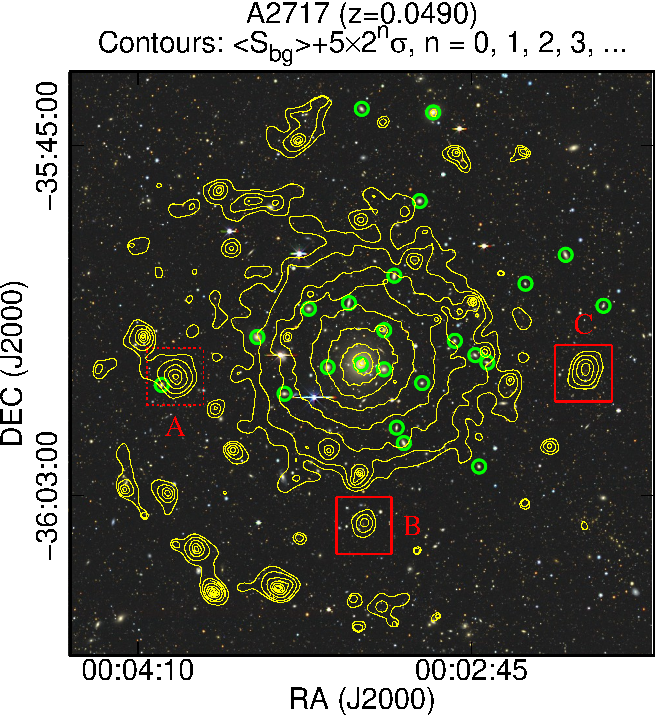}
\end{subfigure}
\hfill
\begin{subfigure}{0.28\textwidth}
\includegraphics[width=\textwidth, angle=0]{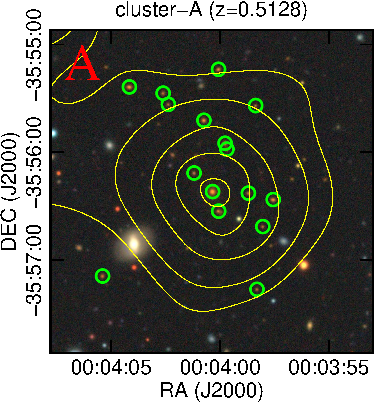}\\[2mm]
\includegraphics[width=\textwidth, angle=0]{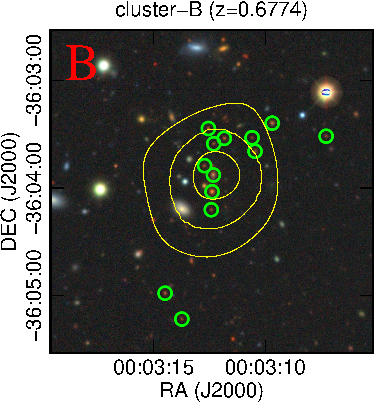}\\[2mm]
\includegraphics[width=\textwidth, angle=0]{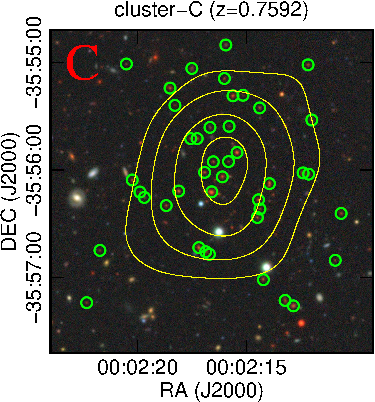}
\end{subfigure}
\caption{An example of X-ray emission detection around
  optically-identified galaxy clusters. Left panel: X-ray-optical
  composite image in the field-of-view of one XMM-Newton observation
  (ObsID = 0145020201). Contours indicate X-ray emissions detected by
  the XMM-Newton, with contour levels labeled at the top of the
  panel. The colorful background displays the DESI DR10 image. The
  central diffuse X-ray structure corresponds to the galaxy cluster
  A2717 at $z=0.0490$, with member galaxies ($M_{*}>10^{10}M_{\odot}$)
  highlighted by green circles, see \citet{wh24} for details. Around
  the A2717, three X-ray sources (``A'', ``B'' and ``C'') are found to
  be associated with optical clusters at much higher redshifts. While
  X-ray cluster ``A'' (R.A. = 1.00109, Dec. = -35.94789) was
  previously reported by \citet{kcs+21}, cluster ``B'' (R.A. =
  0.80111, Dec. = -36.07524) and ``C'' (R.A. = 0.56890, Dec. =
  -35.94470) are newly-identified from X-ray images. The field size of
  the left panel is $30'\times30'$, while the three red squares each
  have a size of $3'\times3'$. Right panels: Zoomed-in views of the
  three high-redshift clusters. Green circles mark the identified
  member galaxies ($M_{*}>10^{10}M_{\odot}$) of each cluster. Panel
  dimensions match the red squares in the left panel.}
\label{fig1}
\end{figure*}

\begin{table*}
\begin{center}
\caption{Parameters for 1,490 newly-identified X-ray clusters (see
  \url{http://zmtt.bao.ac.cn/galaxy\_clusters/} for the full table).}
\tabcolsep=3pt 
\footnotesize
\begin{tabular}{crrccccccccc}
\hline
  \multicolumn{1}{c}{Name}  &\multicolumn{1}{c}{R.A.}  &\multicolumn{1}{c}{Dec.} & $z$  &  \multicolumn{1}{c}{ObsID} &  \multicolumn{1}{c}{$\eta_{300\rm kpc}*$} & $F_{300\rm kpc}$ & \multicolumn{1}{c}{$L_{300\rm kpc}$} &\multicolumn{1}{c}{$r_{500}$}  &\multicolumn{1}{c}{$L_{500}$}  &  \multicolumn{1}{c}{$M_{500}$}   &  \multicolumn{1}{c}{Ref.}\\
  &\multicolumn{1}{c}{(J2000)}& \multicolumn{1}{c}{(J2000)} & & &\multicolumn{1}{c}{($\rm 10^{-3}cnt/s$)} &\multicolumn{1}{c}{($\rm 10^{-14}erg/s/cm^{2}$)} &\multicolumn{1}{c}{($\rm 10^{43}erg/s$)} &\multicolumn{1}{c}{(Mpc)} &\multicolumn{1}{c}{($\rm 10^{43}erg/s$)} &\multicolumn{1}{c}{($10^{14}M_{\odot}$)} &\\
  \multicolumn{1}{c}{(1)} &\multicolumn{1}{c}{(2)}  &\multicolumn{1}{c}{(3)} & \multicolumn{1}{c}{(4)}   &  \multicolumn{1}{c}{(5)} & \multicolumn{1}{c}{(6)} & \multicolumn{1}{c}{(7)} & \multicolumn{1}{c}{(8)} & \multicolumn{1}{c}{(9)} & \multicolumn{1}{c}{(10)} & \multicolumn{1}{c}{(11)} & \multicolumn{1}{c}{(12)} \\
  \hline                    
J0000$-$6020 &0.03709 &$-$60.33922 &0.79 &0862640901 & 4.5$\pm$0.2 & 3.55$\pm$0.18 & 9.58$\pm$0.48  &0.57 &11.17$\pm$0.56 &1.62$\pm$0.18 &1 \\       
J0000$-$6024 &0.04707 &$-$60.41596 &0.19 &0862640901 & 8.7$\pm$0.4 & 7.43$\pm$0.32 & 0.72$\pm$0.03  &0.48 & 0.79$\pm$0.03 &0.67$\pm$0.06 &1 \\       
J0000$-$3935 &0.07040 &$-$39.59931 &0.35 &0864053501 & 7.9$\pm$0.7 & 5.95$\pm$0.53 & 2.40$\pm$0.22  &0.56 & 2.78$\pm$0.25 &1.21$\pm$0.24 &1 \\       
J0001$-$3446 &0.31549 &$-$34.76810 &0.73 &0655300101 & 2.1$\pm$0.2 & 1.53$\pm$0.12 & 2.69$\pm$0.21  &0.46 & 2.94$\pm$0.23 &0.75$\pm$0.13 &1 \\       
J0001$-$5440 &0.40911 &$-$54.67283 &0.90 &0862641401 & 6.7$\pm$0.4 & 4.87$\pm$0.28 &20.91$\pm$1.19  &0.62 &25.10$\pm$1.43 &2.32$\pm$0.29 &1 \\       
J0002$-$3556 &0.56890 &$-$35.94470 &0.76 &0145020201 & 5.2$\pm$0.2 & 3.59$\pm$0.16 & 8.79$\pm$0.38  &0.56 &10.23$\pm$0.44 &1.59$\pm$0.15 &1 \\       
J0003$-$3604 &0.80111 &$-$36.07524 &0.68 &0145020201 & 2.9$\pm$0.1 & 2.24$\pm$0.10 & 3.70$\pm$0.16  &0.50 & 4.14$\pm$0.18 &1.00$\pm$0.10 &1 \\       
J0003$-$0149 &0.86915 & $-$1.82191 &0.45 &0653290101 & 2.7$\pm$0.2 & 2.38$\pm$0.15 & 1.64$\pm$0.11  &0.50 & 1.84$\pm$0.12 &0.81$\pm$0.11 &1 \\       
J0003$+$1001 &0.93253 &   10.02331 &0.37 &0881900801 &56.2$\pm$1.1 &38.98$\pm$0.78 &22.36$\pm$0.45  &0.91 &32.02$\pm$0.64 &5.59$\pm$0.24 &1 \\       
J0013$-$3022 &3.40566 &$-$30.37555 &1.17 &0743850101 & 1.2$\pm$0.1 & 0.71$\pm$0.05 & 3.27$\pm$0.22  &0.37 & 3.44$\pm$0.23 &0.45$\pm$0.07 &1 \\       
\hline                  
\end{tabular}
\end{center}
{Columns: (1) Cluster name. (2 - 4) Right ascension (in J2000),
  declination (in J2000) and redshift of the cluster. These parameters
  are taken from the literature listed in the last column. If the
  cluster belongs to Sample 1, these parameters are obtained from
  Sample 1; if not, the parameters are derived from Sample 2. We
  retain the precision of the redshift to two decimal places. (5)
  Observation ID of the XMM-Newton satellite. (6) Net count rate in
  the [0.5 - 2.0] keV band, within a 300 kpc radius centered on the
  flux-weighted center. The data uncertainty is calculated from the
  Poisson error of net photon counts. If the count rate is marked with
  ``*'', it means that the count rate of this cluster may be
  underestimated due to a small part of the cluster lying beyond the
  CCD coverage. Consequently, parameters in columns 7 - 11 may also be
  underestimated. (7 - 8) Estimated X-ray flux and luminosity in the
  [0.5 - 2.0] keV band, within a 300 kpc radius.  (9 - 11) The radius
  $r_{500}$, the luminosity in [0.5-2.0] keV, and total mass estimated
  within this radius. (12) Cluster sample(s) that include this
  cluster, see Sections~\ref{searchnew} for details. Reference for the
  cluster sample is denoted as (1) = WH24 \citep{wh24}; (2) = WHY18
  \citep{why18}.}
\label{tab1}      
\end{table*}

\subsection{Searching for known X-ray galaxy clusters from XMM-Newton images}
\label{searchknown}

To extract as many galaxy clusters as possible while ensuring their
reliability, we search for clusters from XMM-Newton X-ray images using
the coordinates supplied by published cluster catalogues. Firstly, we
search for known X-ray clusters with positions supplied by three X-ray
cluster samples: the XMM Cluster Archive Super Survey sample
\citep[X-CLASS,][]{kcs+21}, the first eROSITA All-Sky Survey cluster
sample \citep[eRASS1,][]{blk+24} and the second release of the
Meta-Catalogue of X-ray detected Clusters of galaxies
\citep[MCXC-II,][]{sap+24}. In the X-CLASS sample, \citet{kcs+21}
detected 1,646 X-ray galaxy clusters from 4,176 archival observations
of the XMM-Newton satellite. Based on data from the first six months
of the western Galactic hemisphere of the eROSITA All-Sky Survey,
\citet{blk+24} identified 12,247 X-ray galaxy clusters. The eRASS1
cluster catalogue now stands as the largest X-ray selected sample of
galaxy clusters. \citet{pap+11} constructed the MCXC catalogue by
consolidating publicly available cluster samples
\citep[e.g.,][]{eeb+98,eea+00,eeh01,bvh+00,bsg+04} derived from the
ROSAT All-Sky Survey data. Recently, \citet{sap+24} published the
second release of the MCXC catalogue with a total of 2,221 galaxy
clusters.

For each processed X-ray image, we search for X-ray emission excesses
within a circular region of 1-arcminute radius centered on the
positions provided by the three aforementioned X-ray cluster
catalogues. All matched emission excesses are regarded as X-ray
cluster candidates. Note that the X-ray emission around cluster
positions may originate from unrelated foreground or background
sources. We further carefully examine the candidates by overlaying
X-ray images onto DESI optical
images\footnote{https://www.legacysurvey.org/viewer}, as illustrated
in Fig.~\ref{fig1}. Candidates whose X-ray peaks coincide with
unrelated optical sources, such as AGNs or stars, are discarded.  We
also correct a small number of optical-X-ray counterpart mismatches
resulting from the line-of-sight superposition of multiple optically
identified clusters. Finally, we obtain 1913 known X-ray clusters from
the XMM-Newton images. Their parameters are listed in Table A1.

\subsection{Searching for previously unrecognized X-ray clusters}
\label{searchnew}

In addition to known X-ray clusters, the primary goal of this study is
to detect as many previously unidentified X-ray clusters as possible
from XMM-Newton images. Our method first involves searching for excess
X-ray emission around the positions of known optical clusters, and
then verifying the associations between the X-ray emission and the
corresponding optical clusters. We use two optical cluster catalogues
-- the WHY18 catalogue \citep{why18} and the WH24 catalogue
\citep{wh24} -- to search for new X-ray galaxy clusters. Based on the
photometric data from the 2MASS, WISE and SuperCOSMOS, \citet{why18}
published the WHY18 catalogue, which covers a sky area of about 28,000
square degrees and contains 47,600 galaxy clusters. Among these,
26,125 clusters were newly identified, most of which lie outside the
sky coverage of Sloan Digital Sky Survey (SDSS). The detection rate is
$\sim90\%$ for clusters with $M_{500}>3\times10^{14}M_{\odot}$ within
the redshift range of $0.025<z<0.3$, and the false detection rate for
the entire sample is less than $5\%$. By utilizing data from the DESI
Legacy Imaging Surveys and available spectroscopic redshifts,
\citet{wh24} identified over 1.58 million galaxy clusters across a sky
region around 24,000 square degrees, which currently represents the
largest sample of galaxy clusters, including 877,806 clusters
identified for the first time. The WH24 catalogue demonstrates high
completeness and purity for low-redshift and rich clusters. For
instance, it achieves a detection rate exceeding $90\%$ for clusters
with masses over $10^{14}M_{\odot}$ when cross-matched with
established catalogues \citep[see figure 10 in][]{wh24}. In the COSMOS
field, comparison with known catalogues reveals that 141 out of 145
clusters identified at $z<0.9$ had been reported, and the rest four
detections are also likely genuine clusters, indicating a high purity.

Similar to the search for known X-ray clusters, we detect previously
unrecognized X-ray clusters around positions supplied by the WHY18 and
WH24 catalogues within a 1-arcminute radius. To enhance detection
reliability, for X-ray excesses solely associated with optical
clusters, we accept them as the X-ray emission of the ICM if they meet
two additional conditions: (1) the detected net photon count rate
exceeds 0.001 cnt/s; (2) the extension of the $5\sigma$ contour for
the detected X-ray emission exceeds 100 kpc. Following the procedure
in Section~\ref{searchknown}, we check the X-ray-optical superimposed
images for every candidate to remove misassociations. Finally, we
identify 1490 galaxy clusters newly identified in the X-ray band. The
parameters for these clusters are listed in Table 1.

These newly identified X-ray clusters were not included in previous
X-CLASS, eRASS1, and MCXC-II catalogues, which may be attributed to
several reasons. First, as survey-based cluster catalogues, the
X-CLASS, MCXC-II and eRASS1 samples may miss some sources when
pursuing high purity, depth homogeneity, or availability of ancillary
data \citep[e.g.,][]{blk+24}, metrics critical for some statistical
studies. Second, compared to RASS and eRASS data, the XMM-Newton
observations we used offer deeper exposures (>10 ks) and higher
spatial resolution, significantly enhancing the identification of
high-redshift or faint X-ray galaxy clusters. The flux limit of the
MCXC-II and eRASS1 samples are approximately
$2\times10^{-14}\rm~ergs~cm^{-2}~s^{-1}$ \citep{sap+24} and
$4\times10^{-14}\rm~ergs~cm^{-2}~s^{-1}$ \citep{blk+24}, while the
flux limit of this work is about
$7\times10^{-15}\rm~ergs~cm^{-2}~s^{-1}$. In Fig.~\ref{fig1}, we show
three high-redshift X-ray clusters (marked as ``A'', ``B'' and ``C'')
found around the known cluster A2717 ($z=0.0490$). While this region
is covered by eRASS1 data -- where A2717 is unambiguously detected --
the three high-redshift clusters remain undetectable due to
insufficient exposure. Third, the optical data used in our work differ
from those employed by X-CLASS, eRASS1, and MCXC-II for validation,
and such a difference may also contribute to variations between the
samples. Lastly, the X-CLASS sample incorporates XMM-Newton archival
data through August 2015, this work utilizes data extending through
May 2025.

Evaluating the completeness of our sample is challenging, primarily
because the XMM-Newton observations used do not cover continuous sky
regions, and exposure times and observational configurations vary
across individual observations. Additionally, the optical and X-ray
cluster catalogues used in our identification process exhibit diverse
and complex selection functions. Specifically, the MCXC-II sample
\citep{sap+24} is a composite of multiple heterogeneous samples,
making it difficult to assess the overall completeness of the merged
catalogue. Nevertheless, we can construct a complete subsample of
nearby massive galaxy clusters from our dataset, as nearly all such
clusters have XMM-Newton observations. For instance, \citet{yh20}
successfully built a volume-complete subsample of 125 galaxy clusters
using Chandra archival data.

The final purity of our sample primarily depends on two factors: (i)
the inherent purity of the cluster samples used, and (ii) the purity
introduced through our identification procedure. The optical cluster
samples employed demonstrate high purity at low redshifts or high
signal-to-noise ratios, see \citet{why18} and \citet{wh24} for
details. Meanwhile, the eRASS1 catalogue exhibits an aggregate purity
of $\sim86\%$, enhanceable to nearly $100\%$ through additional
selection constraints \citep[see Table 1 in][]{blk+24}. All three
X-ray samples used here implemented multiwavelength cross-matching to
improve purity. When identifying X-ray emission from galaxy clusters
around specific coordinates, misclassification of unrelated
foreground/background sources (e.g., AGNs) may occur. \citet{wsf+09}
processed archival data from XMM-Newton's first seven years, detecting
18,804 extended sources across $\rm\sim360~deg^{2}$, corresponding to
a density of $\rm\rho\sim0.015~arcmin^{-2}$. Subsequently,
\citet{rww+16} and \citet{wct+20} analyzed 15/20 years of XMM-Newton
data, finding densities of $0.018$ and $\rm0.019~arcmin^{-2}$
respectively. These densities imply a probability of $\pi r_{\rm
  1arcmin}^{2}\times\rho\lesssim0.06$ for a random, unrelated X-ray
source within our 1-arcminute search radius. Consequently, the
estimated contamination is $\sim6\%$, yielding $\sim94\%$ sample
purity. Moreover, as detailed in Section 2.2, we performed careful
visual inspections of optical/X-ray composite images to eliminate
potential misidentifications, thereby further enhancing the sample
purity.

\subsection{Parameter estimation and comparing to previous samples}
\label{paraestimate}

Before estimating cluster parameters, point sources should be
subtracted. We use the \texttt{wavdetect} tool of the CIAO system to
detect point sources and the \texttt{dmfilth} tool to subtract them
while filling the ``holes'' with ambient brightness. The process of
point-source subtraction requires careful manual inspection of
combined X-ray-optical data, as we need to ensure all point sources
(e.g., AGNs and stars) are removed while avoiding mistakenly
subtracting cluster cool-cores or high-redshift clusters as point
sources. Following our previous works \citep[e.g.,][]{yh20,yhw22}, the
point-source-subtracted images are smoothed to a scale of 30 kpc to
avoid the bias that the image pixels represent different physical
scales for sources at varying redshifts.

Although the radius $r_{500}$, within which the average matter density
is 500 times the critical density of the universe, is widely used in
cluster research, many clusters are detected only in smaller regions
due to insufficient exposure time for high-redshift clusters or
limitations in CCD coverage for low-redshift clusters. The luminosity
within 300 kpc is well correlated with cluster mass
\citep[e.g.,][]{blk+24}. Thus, we measure the vignetting corrected
count rate within a circular region of 300 kpc radius ($\eta_{300\rm
  kpc}$) and calculate the cluster luminosity and mass within both 300
kpc and $r_{500}$.

\begin{figure}
\centering
\includegraphics[width=0.37\textwidth, angle=0]{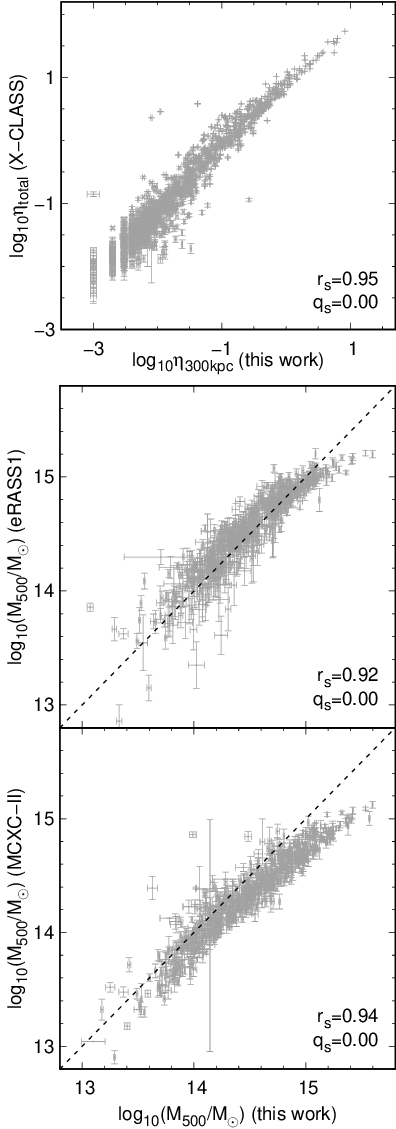}
\caption{Comparison of parameters estimated in this work and
    previous studies. Top-panel: Comparison of count rates from our
    sample ($\eta_{\rm 300kpc}$) and the X-CLASS sample ($\eta_{\rm
      total}$). Note that $\eta_{\rm 300kpc}$ values are calculated
    within a 300 kpc radius, while $\eta_{\rm total}$ from the X-CLASS
    sample are fitted for the entire cluster. Values of the Spearman
    rank-order correlation coefficient $r_{\rm s}$ and correlation
    significance $q_{\rm s}$ are labeled at the lower-right
    corner. Middle-panel: Comparison of cluster masses $M_{500}$ from
    this study with the eRASS1 sample. The dashed line indicates the
    1:1 relation between the X- and Y-axes. Bottom-panel: $M_{500}$
    comparison with the MCXC-II sample.}
\label{fig2}
\end{figure}

The method for parameter estimation adapts approaches from
\cite{xrp+22,xjl+24}. We employ the $\beta$-model for cluster
surface-brightness profile:
\begin{equation}
S_{\rm X}(r) \propto [1+(r/r_{\rm 0})^2]^{-3\beta+0.5},
\label{eq1}
\end{equation} 
where $r_{\rm 0}$ represents the cluster core radius and $\beta$
denotes the profile slope, commonly fixed at 2/3 for standard
clusters. The estimation workflow proceeds as follows:
\begin{itemize}
    \item[1.] The core radius $r_{\rm 0}$ and the radius $r_{\rm 500}$
      are initialized to be 150 kpc and $7r_{\rm 0}$, respectively.
    \item[2.] With the known redshift and assumed radius $r_{\rm
      500}$, the cluster mass within $r_{\rm 500}$ can be calculated
      using $M_{500}=\frac{4}{3}\pi r_{500}^3\times (500
      \rho_{c,~z})$, where $\rho_{c,~z}$ is the critical density at
      cluster redshift.
    \item[3.] The bolometric X-ray luminosity ($L_{\rm bol}$) and
      temperature ($T$) of clusters can be derived from $M_{500}$
      through scaling relations of galaxy clusters. Here, we use the
      scaling relations presented in \citet{rbf+11} as,
      \begin{equation}
        \begin{split}
      & M_{500}={A_1}\cdot L_{\rm bol}[10^{44}{\rm erg/s}]^{B_1}\cdot E(z)^{C_1}\times10^{14}M_{\odot},\\
      & M_{500}={A_2}\cdot T[{\rm keV}]^{B_2}\cdot E(z)^{C_2}\times10^{14}M_{\odot},
        \end{split}
        \label{eq2}
      \end{equation} 
      where the values of [A1, B1, C1] are [1.64, 0.52, -0.90] and the
      values of [A2, B2, C2] are [0.291, 1.62, -1.04], refering to the
      equations 26 and 23 in \citet{rbf+11}. Then, we fit the
      Eq.~\ref{eq2} to eRASS1 clusters, and obtain the best-fitting of
      these six free parameters using the {\tt curve\_fit}
      package\footnote{https://docs.scipy.org/doc/scipy/reference/generated/scipy.optimize.curve\_fit.html. The
      sets} of best-fitting values from the eRASS1 clusters are [A1,
        B1, C1] = [4.18, 0.64, -2.32] and [A2, B2, C2]=[0.33, 2.18,
        1.85]. The latter sample consists of more than 40 times the
      number of clusters compared to that in \citet{rbf+11}. Thus, in
      this work, we adopt the values of [A1, B1, C1] and [A2, B2, C2]
      corresponding to the eRASS1 sample as a more reasonable choice.
    \item[4.] Using the bolometric luminosity ($L_{\rm bol}$), we
      estimate the rest-frame luminosity within the [$0.5-2.0$] keV
      band ($L_{500}$), and the observed flux in the [$0.5-2.0$] keV
      band ($F_{500}$) with the $k$-correction applied
      \citep{hbb+02,br07}. In this step, the APEC model (the
      Astrophysical Plasma Emission Code, \citealt{sbl+01,fjs+12}) is
      adopted. The parameters used include the neutral hydrogen
      absorption obtained from the HI4PI survey \citep{HI4PI}, a fixed
      metallicity of $0.3~Z_\odot$, the cluster temperature ($T$)
      derived from step 3, and the cluster redshift.
    \item[5.] We utilize the
      PIMMS\footnote{https://heasarc.gsfc.nasa.gov/cgi-bin/Tools/w3pimms/w3pimms.pl}
      tool \citep{m93} to convert the flux ($F_{500}$) into the count
      rate in the [$0.5-2.0$] keV ($\eta_{500}$). In this step, the
      instrument XMM-Newton MOS2 with a MEDIUM filter is selected for
      consistency with the data processed by the SAS tool
      \texttt{combimage}.
    \item[6.] Using the typical $\beta$-profile (Eq.~\ref{eq1}) with
      $\beta=2/3$, the expected count rate within 300~kpc ($\eta_{\rm
      300kpc,~est}$) can be estimated with known $r_{500}$ and
      $\eta_{500}$.
    \item[7.] To match the observed and estimated values of $\eta_{\rm
      300kpc}$, steps 2 to 6 are iterated until $\eta_{\rm
      300kpc,~est}=\eta_{\rm 300kpc,~obs}$. Subsequently, parameters
      estimated in $r_{500}$ like luminosity ($L_{500}$) and mass
      ($M_{500}$) are obtained.
    \item[8.] Following the step 7, the rest-frame luminosity within
      300 kpc ($L_{300\rm kpc}$) can be estimated from $\eta_{\rm
        300kpc,~obs}$ using the same APEC model and PIMMS
      settings. All estimated parameters are listed in Table 1.
\end{itemize}

We compare parameters estimated in this work with literature
values. Given that cluster radius $r_{500}$, luminosity $L_{500}$ and
mass $M_{500}$ are correlated \citep[e.g.,][]{app05,crb+07,mae+10}, we
focus on comparing the fundamental parameter $M_{500}$ derived here
with values from previous X-ray samples. Since $M_{500}$ is
unavailable in the X-CLASS sample, we instead compare our count rates
($\eta_{\rm 300kpc}$) with those ($\eta_{\rm total}$) from the X-CLASS
sample, where the former are calculated within a radius of 300 kpc
while the latter are estimated for the full cluster using a fitting
model \citep[see][for details]{kcs+21}. In the top-panel of
Fig.~\ref{fig2}, count rates from our sample and the X-CLASS sample
show a strong correlation despite being measured within different
radii. We calculate the Spearman rank-order correlation coefficient
$r_{\rm s}$ and the significance coefficient $q_{\rm s}$ \citep[see
  definition in ][page 640]{ptv+92} between the $\eta_{\rm 300kpc}$
and $\eta_{\rm total}$. Here, when $r_{\rm s}$ approaches a value of 1
(-1), it indicates a positive (negative) correlation between the two
variables; when $r_{\rm s}$ approaches 0, it indicates no correlation
between them. Furthermore, $q_{\rm s}$ quantifies the significance
level of the correlation: a $q_{\rm s}$ value of 0 denotes a
significant correlation, whereas a notable deviation from 0 (e.g.,
$q_{\rm s}>0.05$) indicates either the absence of a correlation or
that the correlation result is rendered unreliable due to an
inadequate sample size. In the top panel, the results of $r_{\rm
  s}=0.95$ and $q_{\rm s}<10^{-6}$ further confirm a robust
correlation between $\eta_{\rm 300kpc}$ and $\eta_{\rm total}$. The
middle and bottom panels of Fig.~\ref{fig2} demonstrate good
consistency between our $M_{500}$ estimates and those from the eRASS1
and MCXC-II samples, with $r_{\rm s}$ and $q_{\rm s}$ values further
validating this agreement. Nevertheless, systematic offsets exist
between different samples, potentially arising from differences in
instrumentation or parameter calibrations.

\begin{figure*}
\centering
\includegraphics[width=0.8\textwidth, angle=0]{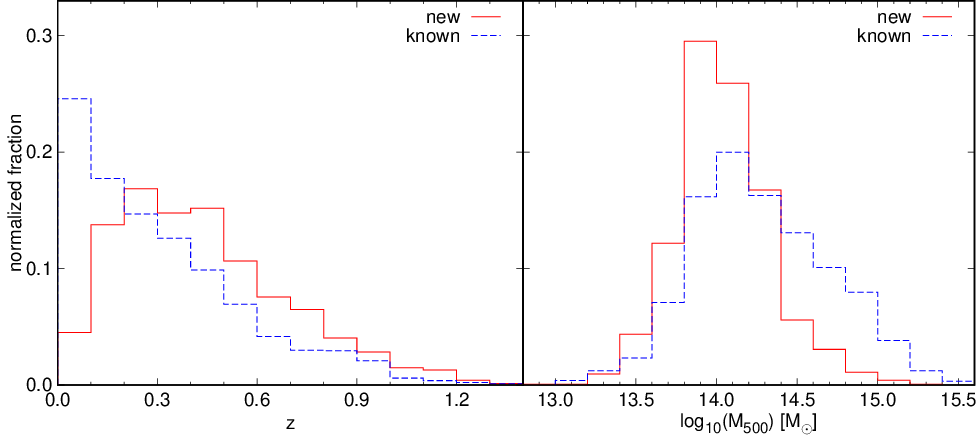}\\[3mm]
\includegraphics[width=0.8\textwidth, angle=0]{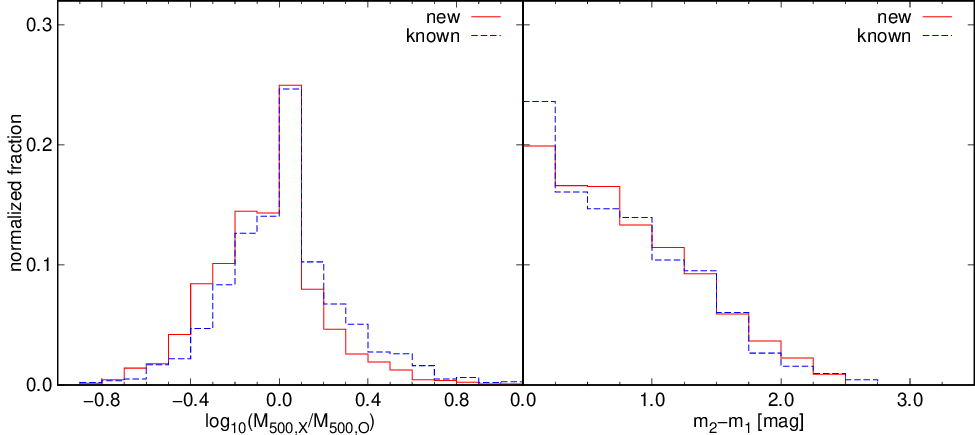}
\caption{The normalized distribution of redshift (upper-left), X-ray
  mass (upper-right), X-ray-to-optical mass ratio (lower-left) and
  magnitude difference between the 1st and 2nd brightest member
  galaxies (lower-right) for clusters in our sample. The ``known''
  subsample (dashed line) denotes clusters previously included in the
  X-CLASS, eRASS1 and MCXC-II samples, while the ``new'' subsample
  (solid line) is for clusters not presented in these three X-ray
  samples.}
\label{fig3}
\end{figure*}

\section{Comparing known and new X-ray clusters}
\label{sect3}

In this section, we investigate the parameter distributions of galaxy
clusters included in our sample and study the distinctive features of
the newly-identified X-ray clusters by comparing them to known X-ray
clusters. As described in Section~\ref{searchknown} and
\ref{searchnew}, the clusters detected in XMM-Newton images are
divided into two subsamples: one contains 1913 clusters previously
included in the X-CLASS, eRASS1 and MCXC-II samples (hereafter the
``known'' subsample), and the other includes 1490 clusters not
presented in these three X-ray samples (hereafter the ``new''
subsample).

In the upper-left panel of Fig.~\ref{fig3}, we show the redshift
distribution of the ``known'' and ``new'' subsamples. The number of
clusters in the known subsample decreases continuously from low to
high redshift, whereas the fraction of clusters in the new subsample
shows a peak at $0.1<z<0.5$. When the redshift exceeds 0.2, the
normalized fraction of galaxy clusters in the new subsample surpasses
that in the known subsample. As stated in Section~\ref{searchnew},
this likely results from the deeper exposures of the XMM-Newton images
compared to RASS/eRASS data. Therefore, we suggest that the subsample
of newly-identified X-ray clusters tends to contain more high-redshift
galaxy clusters than the known subsample.

In the upper-right panel of Fig.~\ref{fig3}, it shows that when the
cluster mass is less than $10^{14}~M_{\odot}$, the fraction of
newly-recognized X-ray clusters exceeds that of known ones. However,
when the mass exceeds $10^{14.5}~M_{\odot}$, this trend
reverses. Thus, compared to the known subsample, the new subsample
tends to contain more low-mass X-ray clusters. From the two upper
panels, we conclude that in the XMM-Newton observed sky, most massive
nearby galaxy clusters have been identified and characterized in
X-ray. Our newly identified X-ray clusters, based on coordinates from
optical catalogues, are typically fainter in X-ray observations. This
is primarily due to the flux limit of our sample is
$\sim7\times10^{-15}\rm~ergs~cm^{-2}~s^{-1}$, which is about 1/3(1/6)
of the MCXC-II(eRASS1) sample. Considering their distance, we are more
likely to discover clusters with higher redshifts or smaller masses.

Previous studies have shown that the total optical luminosity of
cluster member galaxies and the total X-ray luminosity of the ICM are
strongly correlated with cluster mass, making them common mass proxies
for galaxy clusters \citep[e.g.,][]{rb02,wh15}. Since our
newly-identified X-ray clusters were detected based on optical cluster
positions -- a methodology distinct from traditional X-ray cluster
identification, we investigated potential property differences between
the new and known subsamples. In the lower-left panel of
Fig.~\ref{fig3}, we show the distribution of the X-ray-to-optical mass
ratio ($M_{\rm 500,X}/M_{\rm 500,O}$) for both subsamples, with
$M_{\rm 500,O}$ values taken from the WH24 catalogue \citep{wh24}. The
data reveal that the newly-identified X-ray clusters exhibit
systematically lower $M_{\rm 500,X}/M_{\rm 500,O}$ ratios than known
X-ray clusters, indicating deficient X-ray emission relative to their
optical brightness in these systems. Although the $M_{\rm 500,X}$ and
$M_{\rm 500,O}$ presented here were not computed within the strictly
same $r_{500}$ apertures, we emphasize that: (1) identical
mass-calculation methods were applied to both the known and new
subsamples, and (2) optical $M_{\rm 500, O}$ and $r_{500}$
measurements were calibrated to the X-ray reference sample
\citep[see][for details]{wh15,wh24}. Consequently, subtle mass
variations in individual clusters induced by random discrepancies
between optical and X-ray $r_{500}$ determinations are unlikely to
significantly affect systematic differences between these two
subsamples. We therefore suggest that the observed systematic
differences between the new and known subsamples are physical.

The dynamical state is a fundamental characteristic of galaxy
clusters, affecting many cluster properties
\citep[e.g.,][]{yhw15,yhw16,ljv+23,ksj+24}. Various methods are
developed to describe cluster dynamical states, such as those using
the distribution of member galaxies \citep[e.g.,][]{evn+12,wh13,wh24}
or hot gas \citep[e.g.,][]{yhw22,sbb+25,zkk+25}. Since newly-confirmed
X-ray clusters are systematically fainter than known ones,
X-ray-derived dynamical parameters for the two subsamples may exhibit
systematic biases. In merging clusters, multiple brightest cluster
galaxies (BCGs) of progenitors with comparable masses/luminosities
typically coexist, whereas relaxed clusters generally feature dominant
central galaxies whose masses/luminosities exceed satellite galaxies
significantly \citep[e.g.,][]{wh13,wh15b}. We therefore adopt the
$z$-band magnitude difference between the first and second brightest
member galaxies, $m_2-m_1$, as a dynamical probe of galaxy
clusters. In the lower-right panel of Fig.~\ref{fig3}, we compare the
newly-confirmed and known subsamples in dynamical properties and find
no significant difference in the $m_2-m_1$ distributions between the
two subsamples, indicating comparable dynamical properties between
newly-confirmed and known X-ray clusters.

\section{Summary}
\label{sect4}

Considering current status of galaxy cluster identification,
galaxy-based detection methods demonstrates significantly higher
efficiency compared to ICM-based methods, yielding a sample size
disparity of approximately two orders of magnitude between the two
approaches. Traditional X-ray cluster identification typically
involved candidate pre-selection via specific criteria, which
inevitably introduces the omission of faint X-ray clusters. In this
study, we used a new method to search for previously-unrecognized
X-ray clusters from XMM-Newton archival data, utilizing the
coordinates of galaxy clusters provided by large optical galaxy
cluster catalogues. By combining known X-ray clusters, we ultimately
identified 3403 galaxy clusters from XMM-Newton images.

We estimated parameters for galaxy clusters in our sample and compared
them with previous studies, finding overall consistent
results. Comparing newly-confirmed X-ray clusters with
previously-identified ones, we found that the former exhibit
systematically higher redshifts, smaller masses and smaller
X-ray-to-optical mass ratios. No significant difference emerge in the
dynamical properties of the newly-confirmed versus known subsamples.

\section*{Acknowledgements}
We thank the anonymous referee for instructive comments that have
helped us improve the paper. ZSY acknowledges Prof. Junjie Mao and
Ms. Jiejia Liu for helpful discussions on data processing. This work
is partially supported by the National Natural Science Foundation of
China (Grant Nos. 11988101, 12022306, 12203063, 12073036, 12425303),
science research grants from the China Manned Space Project (Grant
Nos. CMS-CSST-2025-A04, CMS-CSST-2025-A03) and the National SKA
Program of China (Grant Nos. 2022SKA0120103, 2020SKA0110100). WX
thanks the National Key R$\&$D Program of China (Grant
No. 2022YFF0503403), the CAS Project for Young Scientists in Basic
Research (No. YSBR-062), and the K.C.Wong Education Foundation for
their support.

This work is based on observations obtained with XMM-Newton, an ESA
science mission with instruments and contributions directly funded by
ESA Member States and the National Aeronautics and Space
Administration (NASA). This research has made use of the NASA/IPAC
Extragalactic Database (NED), which is funded by NASA and operated by
the California Institute of Technology.

The DESI Legacy Imaging Surveys consist of three individual and
complementary projects: the Dark Energy Camera Legacy Survey (DECaLS),
the Beijing-Arizona Sky Survey (BASS), and the Mayall z-band Legacy
Survey (MzLS). DECaLS, BASS, and MzLS together include data obtained,
respectively, at the Blanco telescope, Cerro Tololo Inter-American
Observatory, NSF’s NOIRLab; the Bok telescope, Steward Observatory,
University of Arizona; and the Mayall telescope, Kitt Peak National
Observatory, NOIRLab. NOIRLab is operated by the Association of
Universities for Research in Astronomy (AURA) under a cooperative
agreement with the National Science Foundation. Pipeline processing
and analyses of the data were supported by NOIRLab and the Lawrence
Berkeley National Laboratory (LBNL). Legacy Surveys also uses data
products from the Near-Earth Object Wide-field Infrared Survey
Explorer (NEOWISE), a project of the Jet Propulsion
Laboratory/California Institute of Technology, funded by the National
Aeronautics and Space Administration. Legacy Surveys was supported by:
the Director, Office of Science, Office of High Energy Physics of the
U.S. Department of Energy; the National Energy Research Scientific
Computing Center, a DOE Office of Science User Facility; the
U.S. National Science Foundation, Division of Astronomical Sciences;
the National Astronomical Observatories of China, the Chinese Academy
of Sciences, and the Chinese National Natural Science Foundation. LBNL
is managed by the Regents of the University of California under
contract to the U.S. Department of Energy. The complete
acknowledgments can be found at
https://www.legacysurvey.org/acknowledgment/.

\section*{Data availability}

The data underlying this article will be shared upon reasonable
request to the first author. Full Tables 1 and A1 can be found on the
webpage: http://zmtt.bao.ac.cn/galaxy\_clusters/.

\bibliographystyle{mnras}
\bibliography{ref}
\begin{appendix}
\section{Table for previously-identified X-ray clusters}

\begin{table*}
\begin{center}
\caption{Parameters for 1,913 known X-ray clusters identified from
  XMM-Newton images (see
  \url{http://zmtt.bao.ac.cn/galaxy\_clusters/} for the full table).}  
\tabcolsep=3pt
\footnotesize
\begin{tabular}{crrccccccccc}
\hline
  \multicolumn{1}{c}{Name}  & \multicolumn{1}{c}{R.A.}  &\multicolumn{1}{c}{Dec.} & $z$  &  \multicolumn{1}{c}{ObsID} &  \multicolumn{1}{c}{$\eta_{300\rm kpc}*$} & $F_{300\rm kpc}$ & \multicolumn{1}{c}{$L_{300\rm kpc}$} &\multicolumn{1}{c}{$r_{500}$}  &\multicolumn{1}{c}{$L_{500}$}  &  \multicolumn{1}{c}{$M_{500}$}   &  \multicolumn{1}{c}{Ref.}\\
  & \multicolumn{1}{c}{(J2000)}& \multicolumn{1}{c}{(J2000)} & & &\multicolumn{1}{c}{($\rm 10^{-3}cnt/s$)} &\multicolumn{1}{c}{($\rm 10^{-14}erg/s/cm^{2}$)} &\multicolumn{1}{c}{($\rm 10^{43}erg/s$)} &\multicolumn{1}{c}{(Mpc)} &\multicolumn{1}{c}{($\rm 10^{43}erg/s$)}  &\multicolumn{1}{c}{($10^{14}M_{\odot}$)} &\\
  \multicolumn{1}{c}{(1)} &\multicolumn{1}{c}{(2)}  &\multicolumn{1}{c}{(3)} & \multicolumn{1}{c}{(4)}   &  \multicolumn{1}{c}{(5)} & \multicolumn{1}{c}{(6)} & \multicolumn{1}{c}{(7)} & \multicolumn{1}{c}{(8)} & \multicolumn{1}{c}{(9)} & \multicolumn{1}{c}{(10)} & \multicolumn{1}{c}{(11)} & \multicolumn{1}{c}{(11)}\\
  \hline                    
J0000$+$0816 &0.02931 &  8.27918 &0.04 &0741581501  &250.6$\pm$8.1  &203.65$\pm$6.61 & 0.71$\pm$0.02  &0.53 & 0.81$\pm$0.03 & 0.80$\pm$0.06 &1,5     \\ 
J0001$-$1540 &0.42262 &-15.67913 &0.13 &0204790101  &154.3$\pm$4.8* &112.12$\pm$3.50 & 5.29$\pm$0.17  &0.76 & 6.90$\pm$0.22 & 2.85$\pm$0.19 &2,5     \\ 
J0001$-$5609 &0.48500 &-56.16028 &0.30 &0862640701  & 21.1$\pm$1.0  & 15.18$\pm$0.69 & 4.67$\pm$0.21  &0.66 & 5.74$\pm$0.26 & 2.04$\pm$0.20 &1,4     \\ 
J0002$-$0202 &0.64964 & -2.03948 &0.38 &0653290201  &  6.3$\pm$0.3  &  4.66$\pm$0.23 & 2.40$\pm$0.12  &0.56 & 2.77$\pm$0.14 & 1.15$\pm$0.12 &1,3     \\ 
J0002$-$3439 &0.67803 &-34.65995 &0.13 &0655300101  & 71.3$\pm$1.7* & 52.46$\pm$1.25 & 2.28$\pm$0.05  &0.63 & 2.75$\pm$0.07 & 1.59$\pm$0.08 &1,2,3,4 \\ 
J0002$-$2955 &0.74200 &-29.92380 &0.06 &0041750101  & 16.8$\pm$0.3  & 14.71$\pm$0.25 & 0.14$\pm$0.01  &0.37 & 0.14$\pm$0.01 & 0.25$\pm$0.01 &3       \\ 
J0003$-$0605 &0.79826 & -6.09167 &0.23 &0652010401  &316.1$\pm$1.5  &210.11$\pm$1.00 &38.88$\pm$0.18  &1.13 &63.55$\pm$0.30 &10.41$\pm$0.11 &1,2,3,5 \\ 
J0003$-$3556 &0.80150 &-35.93540 &0.05 &0145020201  &684.2$\pm$1.6  &501.47$\pm$1.18 & 2.87$\pm$0.01  &0.69 & 3.59$\pm$0.01 & 2.06$\pm$0.01 &3,4,5   \\ 
J0003$+$0203 &0.95694 &  2.06650 &0.10 &0201900101  &354.7$\pm$1.8  &255.67$\pm$1.27 & 7.28$\pm$0.04  &0.83 & 9.89$\pm$0.05 & 3.70$\pm$0.04 &1,2,3,5 \\ 
J0004$-$3556 &1.00109 &-35.94789 &0.51 &0145020201  &  8.1$\pm$0.2  &  5.78$\pm$0.17 & 5.74$\pm$0.17  &0.61 & 6.83$\pm$0.20 & 1.72$\pm$0.11 &1,3     \\ 
\hline                  
\end{tabular}
\end{center}
{{\it Notes.} Columns: (1) Cluster name. (2 - 4) Right ascension
  (J2000), declination (J2000) and redshift of the cluster, taken from
  literature listed in the last column. If the cluster is included by
  sample 1, these three parameters are obtained from sample 1; if not,
  the parameters are adopted from sample 2, and so forth up to sample
  5. (5) Observation ID of the XMM-Newton satellite. (6) Net count
  rate ([0.5 - 2.0] keV) within a 300 kpc aperture centered on the
  flux-weighted center. Asterisk (*) flags the cluster with potential
  count rate underestimation due to partial CCD coverage, propagating
  to underestimated parameters in columns 6-10. (7 - 8) X-ray flux and
  luminosity ([0.5 - 2.0] keV) within the 300 kpc aperture.  (9 - 11)
  The radius $r_{500}$, the luminosity ([0.5-2.0] keV) and total mass
  estimated within this radius. (12) Cross-matched samples for this
  cluster, see Sections~\ref{searchknown} and ~\ref{searchnew} for
  details. References are denoted with numbers (1) = WH24
  \citep{wh24}; (2) = WHY18 \citep{why18}; (3) = X-CLASS
  \citep{kcs+21}; (4) = eRASS1 \citep{blk+24}; (5) = MCXC-II
  \citep{sap+24}.}
\label{tabA1}      
\end{table*}

\end{appendix}

\label{lastpage}
\end{document}